\documentclass[showpacs,twocolumn]{revtex4}

\input{epsf}

\usepackage{subfig}
\usepackage{graphicx}
\usepackage{caption}

 \def\frac#1#2{{\textstyle{{#1}\over
{#2}}}} 
\def\lsim{\mathrel{\rlap{\lower4pt\hbox{\hskip1pt$\sim$}}
\raise1pt\hbox{$<$}}}
\def\gsim{\mathrel{\rlap{\lower4pt\hbox{\hskip1pt$\sim$}}
\raise1pt\hbox{$>$}}} \def\sqr#1#2{{\vcenter{\vbox{\hrule
height.#2pt \hbox{\vrule width.#2pt height#1pt \kern#1pt \vrule
width.#2pt} \hrule height.#2pt}}}}
\def\square{\mathchoice\sqr66\sqr66\sqr{2.1}3\sqr{1.5}3}
\def\beq{\begin{equation}} \def\eeq{\end{equation}}
\def\beqa{\begin{eqnarray}} \def\eeqa{\end{eqnarray}}

\long\def\symbolfootnote[#1]#2{\begingroup
\def\thefootnote{\fnsymbol{footnote}}\footnote[#1]{#2}\endgroup}

\begin{document}

\title{On the dynamics of perfect fluids in non-minimally coupled gravity}

\vskip 0.2cm

\author{Orfeu Bertolami\footnote{Also at  Instituto de Plasmas e Fus\~ao Nuclear, Instituto Superior T\'ecnico, Av. Rovisco Pais 1, 1049-001 Lisboa. E-mail: orfeu.bertolami@fc.up.pt}}

\affiliation{
	\vskip 0.1cm
	Departamento de F\'{\i}sica e Astronomia, Faculdade de Ci\^encias, Universidade do Porto, \\
	Rua do Campo Alegre 687, 4169-007 , Porto, Portugal
}

\author{Ant\'onio Martins\footnote{E-mail: antonio.fpn.martins@gmail.com}}

\affiliation{
	\vskip 0.1cm
	Instituto de Plasmas e Fus\~ao Nuclear, Instituto Superior T\'ecnico \\
	Av.\ Rovisco Pais 1, 1049-001 Lisboa, Portugal
}

\vskip 0.2cm

\date{\today}


\begin{abstract}
	
In this work we explore the consequences that a non-minimal coupling between geometry and matter can have on the dynamics of perfect fluids. It is argued that the presence of a static,  axially symmetric pressureless fluid does not imply a Minkowski space-time like as is in General Relativity. This feature can be atributed to a pressure mimicking mechanism related to the non-minimal coupling. The case of a spherically symmetric black hole surrounded by fluid matter is analyzed, and it is shown that under equilibrium conditions the total fluid mass is about twice that of the black hole. Finally, a generalization of the Newtonian potential for a fluid element is proposed and its implications are briefly discussed.

\vskip 0.5cm

\end{abstract}



\maketitle


\section{Introduction}

The recent revival of interest in alternative gravity models, motivated
chiefly by the possibility of explaining the accelerated expansion
of the Universe and the flattening of the rotation curves of galaxies
without the need to introduce dark energy and matter, has led to a
number of studies on the so-called {}``$f\left(R\right)$
theories'', where the linear scalar curvature term of the Einstein-Hilbert
action is replaced by a more general function of the same invariant \cite{fr}. Although these models have attracted
a great deal of attention, it is interesting to take the generalization
a step further and consider a non-minimal coupling between matter
and geometry~\cite{ExF,f2rI,f2rII}, which can be achieved by postulating an action of the
form
\begin{equation}
S=\int\left[{1 \over 2}f_{1}\left(R\right)+\left(1+\lambda f_{2}\left(R\right)\right)\mathcal{L}_{m}\right]\sqrt{-g}dx^{4},
\label{eq:S}
\end{equation}
where $f_{i}\left(R\right)$ are arbitrary functions of the scalar
curvature $R$, $\mathcal{L}_{m}$ is the Lagrangian density of matter,
$g$ is the metric determinant, and $\lambda$ is a coupling constant
that can be used to gauge the contribution of $f_{2}(R)$. The standard
Einstein-Hilbert action is recovered by taking $f_{2}=0$ and $f_{1}\left(R\right)=2\kappa\left(R-2\Lambda\right)$,
where $\kappa=\frac{c^{4}}{16\pi G}$ and $\Lambda$ is the cosmological
constant.

Varying action (\ref{eq:S}) with respect to the metric coefficients
yields the field equations
\begin{equation}
\left(F_{1}+2\lambda F_{2}\mathcal{L}_{m}\right)R_{\mu\nu}-{1 \over 2}f_{1}g_{\mu\nu}=
\label{eq:FE}
\end{equation}
\[
\left(\nabla_{\mu}\nabla_{\nu}-g_{\mu\nu}\square\right)\left(F_{1}+2\lambda F_{2}\mathcal{L}_{m}\right)+\left(1+\lambda f_{2}\right)T_{\mu\nu},\]
where $F_{i}\left(R\right)=f_{i}'\left(R\right)$, and the energy-momentum
tensor is defined, as usual, by
\begin{equation}
T_{\mu\nu}=-{2 \over \sqrt{-g}}{\delta\left(\sqrt{-g}\mathcal{L}_{m}\right) \over \delta g^{\mu\nu}}.
\label{eq:T}
\end{equation}

Among the various interesting features of this model, perhaps the
most striking is the non-conservation law
\begin{equation}
\nabla_{\mu}T^{\mu\nu}={\lambda F_{2} \over 1+\lambda f_{2}}\left(g^{\mu\nu}\mathcal{L}_{m}-T^{\mu\nu}\right)\nabla_{\mu}R.
\label{eq:NCT}
\end{equation}
This work aims to examine the implications of Eq. (\ref{eq:NCT}) for Kerr-like
metrics and for the generalization of the Newtonian potential.

This paper is organized as follows: section \ref{sec: SAS} explores the possibility
of adapting static, axially symmetric metrics to regions of space-time
permeated by a presureless perfect fluid in or near hydrostatic equilibrium,
and discusses the Schwarzschild-like case in some detail; a generalization
of the Newtonian potential is examined in section \ref{sec: GNP}, and its implications
are briefly discussed. In Section \ref{sec: Conc} we present our conclusions.

\section{Axial symmetry in a static pressureless fluid with non-minimal gravitational coupling}
\label{sec: SAS}

Consider a static, axially symmetric system described by a metric
of the form
\begin{equation}
ds^{2}=g_{tt}dt^{2}+g_{rr}dr^{2}+g_{\theta\theta}d\theta^{2}+g_{\phi\phi}d\phi^{2}+2g_{t\phi}dtd\phi,
\label{eq:Metric}
\end{equation}
with $\partial_{t}g_{\mu\nu}=\partial_{\phi}g_{\mu\nu}=0$. Since
we want to study the effects of a non-minimal coupling between matter
and geometry, it is pointless to consider a vacuum situation. As such,
we will admit a matter distribution modeled by a pressureless perfect
fluid in or close to hydrostatic equilibrium. This case is chosen as
it exhibits two attractive features: firstly, it is sufficiently simple
to be treated analytically, while still displaying the consequences
of a non-minimal coupling; secondly, it can be used to model a
number of physically interesting systems.

One can then write
\begin{eqnarray}
T^{\mu\nu}&=&\rho u^{\mu}u^{\nu} \label{eq:TTPerfectFluid}, \\
u^{\mu}&\simeq&\left(u^{t},0,0,0\right), \nonumber
\end{eqnarray}
where $\rho$ is the density, and $\left(u^{t}\right)^{2}=-\left(g_{tt}\right)^{-1}$.
Notice that the last relation implies $g_{tt}<0$, which immediately
disables a Kerr metric identification. Using $\mathcal{L}_{m}=-\rho$ (see Ref. \cite{Lm} for a discussion),
one can readily show that Eq. (\ref{eq:NCT}) imposes the restrictions
\begin{equation}
\partial_{i}\ln\left(-g_{tt}\right)={\partial_{i}\left(-g_{tt}\right) \over \left(-g_{tt}\right)}=-{2\lambda F_{2}\left(R\right) \over 1+\lambda f_{2}\left(R\right)}\partial_{i}R,
\label{eq:gRRelation}
\end{equation}
with $i=r,\theta$. Eq. (\ref{eq:gRRelation}) introduces a rather
surprising functional relation between $g_{tt}$ and $R$. It is interesting
to compare the above result with the equation of hydrostatic equilibrium
(for a fluid with pressure $p$ and density $\rho$) arising from
the spherically symmetric case in General Relativity, often used in
studies of stellar structure~\cite{SS}:
\begin{equation}
{\partial_{r}\left(-g_{tt}\right) \over \left(-g_{tt}\right)}=-{2 \over \rho+p}\partial_{r}p.
\label{eq:GRHEqui}
\end{equation}

The similarity of Eqs. (\ref{eq:gRRelation}) and (\ref{eq:GRHEqui})
suggests the non-minimal coupling acts as an effective pressure. Indeed,
by noting that in the spherically symmetric case $R\equiv R\left(r\right)$,
one has $\frac{\partial}{\partial R}=\frac{\partial r}{\partial R}\frac{\partial}{\partial r}$%
\footnote{In fact, the same relation holds even for the axially symmetric case. See below.%
}, so that the numerator of Eq. (\ref{eq:gRRelation}) reads $2\left(\frac{\partial R}{\partial r}\frac{\partial r}{\partial R}\right)\partial_{r}\left(\lambda f_{2}\right)=2\partial_{r}\left(\lambda f_{2}\right)$,
where we have used $\frac{\partial R}{\partial r}\frac{\partial r}{\partial R}=1$.
If one now considers $\rho\simeq\rho_{0}=constant$ (we shall argue
later that matter in the vicinity of a black hole behaves asymptotically
as $\rho\sim\left(\frac{r_{s}}{r}\right)^{5}$, where $r_{s}=2GM_{BH}$
is the Schwarzschild radius, so this assumption is not too unrealistic
for $r\gg r_{s}$), then $\rho_{0}\partial_{r}\left(\lambda f_{2}\right)=\partial_{r}\left(\lambda\rho_{0}f\right)$,
and one can write Eq. (\ref{eq:gRRelation}) as
\begin{equation}
{\partial_{r}\left(-g_{tt}\right) \over \left(-g_{tt}\right)}=-{2 \over \rho+p_{eff}}\partial_{r}p_{eff},
\label{eq:gRpEff}
\end{equation}
where $p_{eff}=\lambda\rho_{0}f_{2}\left(R\left(r\right)\right)$.

Notice also that if $g_{tt}$ does not depend on $r$ or $\theta$,
neither does $R$ (provided $\lambda F_{2}\neq0$). In particular,
taking the Minkowski case $g_{tt}=-1$ immediately yields $R=constant$,
independently of the remaining metric coefficients (however, remember
we are considering a static situation, so this result does not apply
to the Friedmann-Robertson-Walker metric). Moreover, if $\partial_{r}g_{tt}\neq0$
(or $\partial_{\theta}g_{tt}\neq0$) and $1+\lambda f_{2}\left(R\right)\neq0$,
one can divide both equations to arrive at
\begin{equation}
{\partial_{\theta}g_{tt} \over \partial_{r}g_{tt}}={\partial_{\theta}R \over \partial_{r}R},
\label{eq:gRRelation2}
\end{equation}
which shows an even more straightforward relation between the derivatives
of $g_{tt}$ and $R$.

Before proceeding, two important comments are in order. At first, it
does seem that Eq. (\ref{eq:gRRelation}) is flawed, since it forbids
the Kerr/Schwarzschild case. That apparent problem arises from implicitly
assuming that $\rho\neq0$ when deriving the result. The vaccum case $\rho=0$
yields the trivial relation $0=0$. Second, although it may seem that
Eqs. (\ref{eq:gRRelation}) and (\ref{eq:gRRelation2}) are totally
independent of $\rho$, that is not quite correct. Indeed, the matter
density influences $g_{tt}$, which, in turn, influences $R$. One
can go even further and assume that Einstein's equations hold approximately,
resulting in $R\simeq\rho/2\kappa$, so that Eq. (\ref{eq:gRRelation2})
reads
\begin{equation}
{\partial_{\theta}g_{tt} \over \partial_{r}g_{tt}} \simeq {\partial_{\theta}\rho \over \partial_{r}\rho},
\label{eq:gRoRelation}
\end{equation}
showing a definite dependence on $\rho$. Nonetheless, it should be
stressed that the above equations represent additional constraints
on $g_{tt}$, $R$ and $\rho$.

To continue, we must cast Eq. (\ref{eq:gRRelation}) in a different
form. Firstly, let's show that $\frac{\partial}{\partial R}=\frac{\partial r}{\partial R}\frac{\partial}{\partial r}=\frac{\partial\theta}{\partial R}\frac{\partial}{\partial\theta}$.
Considering $R\equiv R\left(r,\theta\right)$, the chain rule reads
$\frac{\partial}{\partial R}=\frac{\partial r}{\partial R}\frac{\partial}{\partial r}+\frac{\partial\theta}{\partial R}\frac{\partial}{\partial\theta}$.
However, to compute $\frac{\partial r}{\partial R}$ (or $\frac{\partial\theta}{\partial R}$)
one must express $r$ as a function of $R$, which is not possible,
in general, since $R$ also depends on $\theta$. To solve this issue,
define a new variable $\Theta\equiv\Theta\left(r,\theta\right)=\theta$,
so that for sufficiently well-behaved $R$ functions one can write
$r\equiv r\left(R,\Theta\right)$ and $\theta=\theta\left(R,\Theta\right)=\Theta$.
This implies $\frac{\partial\theta}{\partial R}=0$, proving the first
equality. The second can be proved using the same procedure.

By making use of the result derived above, one can now write
\begin{eqnarray}
-{2\lambda F_{2}\left(R\right) \over 1+\lambda f_{2}\left(R\right)}\partial_{i}R&=&-2{\partial_{i}\left(1+\lambda f_{2}\right) \over 1+\lambda f_{2}} \nonumber \\
&=&-2\partial_{i}\left[\ln\left(1+\lambda f_{2}\right)\right],
\label{eq:LogRHS}
\end{eqnarray}
where, as before, we have used $\frac{\partial R}{\partial r}\frac{\partial r}{\partial R}=\frac{\partial R}{\partial\theta}\frac{\partial\theta}{\partial R}=1$.
Inserting Eq. (\ref{eq:LogRHS}) into Eq. (\ref{eq:gRRelation}) yields
\begin{equation}
\partial_{i}\left[-g_{tt}\left(1+\lambda f_{2}\right)^{2}\right]=0\,\,\,\,\,\,\,\, i=r,\theta
\label{eq:gttCons}
\end{equation}
which finally leads to
\begin{equation}
1+\lambda f_{2}\left(R\left(r,\theta\right)\right)={C \over \sqrt{-g_{tt}}},
\label{eq:CouplegttRelation}
\end{equation}
where $C$ is an integration constant. Note that $g_{tt}$ must be
a function of $\lambda$ satisfying $g_{tt}\left(\lambda=0\right)=-1$,
which fixes $C=1$. Although Eq. (\ref{eq:CouplegttRelation}) does not
specify how the coupling depends on $R$, it determines how the function
$f_{2}$ varies with $r$ and $\theta$, provided one knows the functional
form of $g_{tt}$, an information that is potentially more useful.
As an example, it immediately follows that $f_{2}\rightarrow\infty$
as $g_{tt}\rightarrow0$. Moreover, since $g_{tt}$ is related to the
Newtonian potential $\Phi$, one expects that the strength of
the coupling is associated with the motion of test particles.

\subsection{Spherical symmetry}
\label{subsec: A}

Having discussed the model in a general axially symmetric framework,
we are now in conditions to add up further assumptions so that it is possible to evaluate
its applicability to situations of physical interest.

Spherical symmetry is perhaps the simplest case to consider.
In that case, we immediately have the identities
\begin{eqnarray}
g_{t\phi}&=&0, \\
g_{\phi\phi}&=&r^{2}\sin\left(\theta\right), \\
g_{\theta\theta}&=&r^{2}, \\
\partial_{\theta}g_{\mu\nu}&=&0.
\end{eqnarray}
Furthermore, we can assume that $g_{tt}$ and $g_{rr}$ are functions
of $r$ only. Although these restrictions greatly simplify calculations,
Eq. (\ref{eq:FE}) still seem to be too involved to allow for a straightforward
mathematical treatment, even considering the general relativistic
choice $f_{1}=2\kappa R$ for the geometric part of the Lagrangian.
As such, we will try to pursue a different approach here, motivated
chiefly by comparisons with GR. Though less rigorous, it allows establishing some tentative results, possibly shedding some light
into the physical meaning of the non-minimal coupling.

Consider the case of a spherical black hole of mass $M_{BH}$ surrounded
by a pressureless perfect fluid of density $\rho$. We could argue
that the space-time structure in the vicinity of a BH is dominated by
its presence, and try to describe it using the familiar metric $g_{tt}=-1+\frac{r_{s}}{r}$,
$g_{rr}=-1/g_{tt}$, and $g_{\theta\theta}$,$g_{\phi\phi}$ as above, where $r_{s}=2GM_{BH}$
is the usual Schwarzschild radius.
However, this choice has $g_{tt}>0$ for some $r$, violating the
condition imposed upon $g_{tt}$ at the very inception of this model.
Naturally, there are a number of different ways to tackle this issue.
A rather straightforward alternative is to add a constant
$r_{c}\geq r_{s}^{*}$ to the denominator, resulting in $g_{tt}=-1+\frac{r_{s}^{*}}{r+r_{c}}$
and $1+\lambda f_{2}=\sqrt{1+\frac{r_{s}^{*}}{r+r_{c}-r_{s}^{*}}}$.
Here $r_{s}^{*}=\alpha r_{s}$, where $\alpha\simeq3$. This modification stems from the
fact that for $r\gg r_{c}$ one recovers a spherically symmetric vaccum
metric, i.e., $g_{tt}\rightarrow-1+\frac{r_{s}^{*}}{r}$. However,
one must now take into account the mass of fluid present near the
black hole, so that $r_{s}=2GM_{BH}\rightarrow r_{s}^{*}=2GM_{total}$,
where $M_{total}=M_{BH}+M_{f}=\alpha M_{BH}$, and $M_{f}$ is the
fluid's mass. We shall show later that, in first approximation, $M_{f}\simeq2M_{BH}$,
so that $\alpha\simeq3$. Finally, notice that $r_{c}$ controls the
onset of a vaccum metric. Again, we shall show that about 90\% of
the fluid's mass is concentrated inside a sphere of radius $3r_{s}^{*}$,
suggesting that we should take $r_{c}=r_{s}^{*}$. Although this choice makes
the coupling singular at $r=0$, this is not much of a problem, as the curvature 
scalar is also singular there, and it could be taken $r_{c}=r_{s}^{*}+\epsilon$ and later assume $\epsilon\rightarrow0$.
This leads to~%
\footnote{There is a minor issue we have been ignoring so far: as pointed out
earlier, $g_{tt}$ must depend on $\lambda$, so that one recovers
$g_{tt}=-1$ when $\lambda\rightarrow0$. Making the substitution
$r_{s}^{*}\rightarrow\lambda r_{s}^{*}$ is one way to deal with this
problem. However, if $\lambda\neq0$, it can be ``absorbed'' into
$f_{2}$, which is the same as taking $\lambda=1$, so that no problem
arises whatsoever. We will assume that $\lambda\neq0$ throughout the rest
of this paper.%
}
\begin{equation}
g_{tt}=-1+{r_{s}^{*} \over r+r_{s}^{*}} \label{eq:gttCpl},
\end{equation}
and hence
\begin{equation}
1+\lambda f_{2}=\sqrt{1+{r_{s}^{*} \over r}}. \label{eq:CplExplicit}
\end{equation}

Having fixed $g_{tt}$, the only relevant quantity still amiss is
$g_{rr}$. Bearing in mind the discussion presented above, one expects
that sufficiently far away from the black hole a Schwarzschild metric
should reasonably describe the system (provided $f_{1}=2\kappa R$,
an assumption we are considering through the rest of this work).
This motivates the Ansatz $g_{rr}=-1/g_{tt}$, or
\begin{equation}
g_{rr}={1 \over 1-{r_{s}^{*} \over r+r_{s}^{*}}}.
\label{eq:grr}
\end{equation}
It may be that both $g_{tt}$ and $g_{rr}$ deviate considerably from
an exact solution at some points. However, they should hold reasonably
well in the outer region, and we expect that the added constant may approximately
reproduce the dynamics of the inner region too. Moreover, the assumed simple
forms allow for a straightforward calculation of the curvature
scalar
\begin{equation}
R={2r_{s}^{*3} \over r^{2}\left(r+r_{s}^{*}\right)^{3}}.
\label{eq:R}
\end{equation}
As previously pointed out, there is a singularity at $r=0$, and the horizon condition $g_{tt}(r_{H})=0$ yields $r_{H}=0$, 
that is, the event horizon coincides with the singularity.

\begin{figure}

\centering

\includegraphics{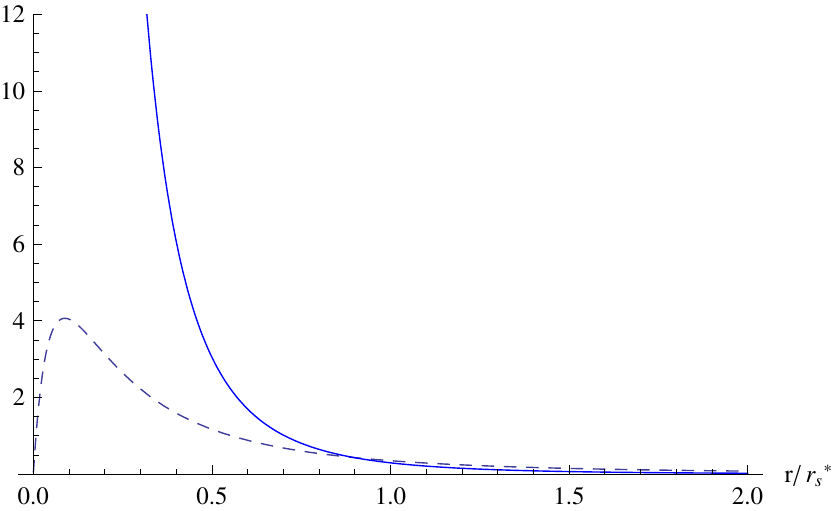}

\caption{Contribution of the various terms on the RHS of Eq. (\ref{eq:Trace})
as a function of $r/r_{s}^{*}$: the Laplacian one $\frac{3}{\kappa}\square\lambda F_{2}\mathcal{L}_{m}$, in dashed line, and the remaining trace term $-\frac{1}{2\kappa}\left(1+\lambda f_{2}\right)T$, in full line. Although they are approximately equal for $r>r_{s}^{*}$, the
trace term clearly dominates in the inner region, where most of the
matter is concentrated.}

\end{figure}
Before proceeding to an explicit calculation of $\rho$, it is instructive
to analyze the physical consequences of a non-minimal coupling.

The trace of Eq. (\ref{eq:FE}) (for $f_{1}=2\kappa R$) reads
\begin{equation}
\left(1-{\lambda \over \kappa}F_{2}\mathcal{L}_{m}\right)R={3 \over \kappa}\square\left(\lambda F_{2}\mathcal{L}_{m}\right)-{1 \over 2\kappa}\left(1+\lambda f_{2}\right)T.
\label{eq:Trace}
\end{equation}
If one neglects the terms containing derivatives of the coupling,
Eq. (\ref{eq:Trace}) can be cast in the approximate form
\begin{equation}
R\simeq{1 \over 2\kappa}\left(1+\lambda f_{2}\right)\rho.
\label{eq:RroRelation}
\end{equation}
Consider now that the non-minimal coupling (NMC) does not disturb the
metric significantly, so that we may take $R^{GR}\simeq R^{NMC}$,
where $R^{GR}$ and $R^{NMC}$ stand for the curvature scalar of General
Relativity (GR) and NMC theory, respectively. Dividing Eq. (\ref{eq:RroRelation})
by the corresponding GR relation leads to
\begin{equation}
{\rho^{GR} \over \rho^{NMC}}=1+\lambda f_{2}.
\label{eq:GR-NMC}
\end{equation}
Eq. (\ref{eq:GR-NMC}) suggests the following physical picture: if
at some point $1+\lambda f_{2}>1$, then $\rho^{GR}>\rho^{NMC}$,
meaning GR's space-time can accommodate a greater density of matter
when compared to the NMC case. From an energetic point of view, and
recalling that we are dealing with a static equilibrium situation, the
amount of matter that minimizes the potential is inferior in the NMC
scenario. The contrary holds if $1+\lambda f_{2}<1$. To summarize,
the NMC is energetically favorable (in the sense that it accommodates
more matter than GR in equilibrium conditions) if $1+\lambda f_{2}<1$,
and unfavorable in case $1+\lambda f_{2}>1$. From Eq. (\ref{eq:CplExplicit})
one immediately has $1+\lambda f_{2}>1$ for all $r$, and indeed
we shall show that $M_{f}^{GR}\gg M_{f}^{NMC}$, reinforcing the physical
reasoning presented above. Finally, note that although we have considered
an example with unfavorable coupling, it is not difficult to envisage
a scenario where $-g_{tt}>1$ and hence $1+\lambda f_{2}<1$. However,
it is not possible to recover the asymptotic Schwarzschild vacuum
in that case, making it a rather odd choice.%

If one inserts Eqs. (\ref{eq:gttCpl}) and (\ref{eq:R}) into (\ref{eq:Trace})
and neglects the Laplacian term, the fluid's density can be easily
computed,
\begin{equation}
\rho\left(r\right)={\rho_{cp} \over \left({r \over r_{s}^{*}}\right)^{3/2}\left(1+{r \over r_{s}^{*}}\right)^{7/2}}\left(1+{r_{s}^{*} \over 5r+r_{s}^{*}}\right),
\label{eq:ro}
\end{equation}
where $\rho_{cp}=\frac{4\kappa}{r_{s}^{*2}}$. In order to justify neglecting
the term $\frac{3}{\kappa}\square\left(\lambda F_{2}\mathcal{L}_{m}\right)$,
we estimate its contribution. From Fig. (1), it is immediately
clear that in the region $r<r_{s}^{*}$, where most of the matter
is concentrated~%
\footnote{Take, for example, the ratio $\rho\left(r_{s}^{*}/2\right)/\rho\left(r_{s}^{*}\right)\simeq8.5$%
}, the term $-\frac{1}{2\kappa}\left(1+\lambda f_{2}\right)T$ dominates.
As such, even though they are approximately equal for $r>r_{s}^{*}$,
the contribution of the Laplacian term is minute, and, in first
approximation, negligible.

Eq. (\ref{eq:ro}) can be exactly integrated to yield the mass of
fluid inside a sphere of radius $r$. However, the resulting expression
is quite cumbersome and not particularly enlightening. In order to
cast the equation in a slightly more manageable form, one may note
that the last term, $1+\frac{r_{s}^{*}}{5r+r_{s}^{*}}$, is dimensionless
and bounded between 1 and 2. This suggests the tentative substitution
$1+\frac{r_{s}^{*}}{5r+r_{s}^{*}}\rightarrow\beta$, where $\beta\in\left]1,2\right[$
is a constant that can be fixed later by comparison with the exact
solution. Performing this replacement, Eq. (\ref{eq:ro}) reads approximately
\begin{equation}
\rho_{\beta}\left(r\right)\simeq\beta{\rho_{cp} \over \left({r \over r_{s}^{*}}\right)^{3/2}\left(1+{r \over r_{s}^{*}}\right)^{7/2}},
\label{eq:robeta}
\end{equation}
where the subscript $\beta$ was added to differentiate it from the
exact case. The interesting feature is that one has now written the
density in the form of a well-known profile type, specifically the
generalized spherical cusped profile~\cite{GCPI,GCPII,GCPIII,GCPIV}, having density scale $\beta\frac{4\kappa}{r_{s}^{*2}}$
and typical length $r_{s}^{*}$. Integrating the result over a sphere
of radius $r$, we find that the mass of fluid enclosed in such volume
to be
\begin{equation}
M_{f}\left(r\right)=\beta{32\pi\kappa r_{s}^{*} \over 15}{r^{3/2}\left(2r+5r_{s}^{*}\right) \over \left(r+r_{s}^{*}\right)^{5/2}},
\label{eq:Mr}
\end{equation}
and taking the limit $r\rightarrow+\infty$, the total fluid mass
reads
\begin{equation}
M_{f}=\beta{64\pi\kappa r_{s}^{*} \over 15}=\beta{8\alpha \over 15}M_{BH},
\label{eq:MfMBH}
\end{equation}
where we have used $\kappa=\frac{1}{16\pi G}=\frac{M_{BH}}{8\pi r_{s}}$.
Inserting Eq. (\ref{eq:MfMBH}) into Eq. (\ref{eq:Mr}), it can be readily
shown that $M_{f}\left(r_{s}^{*}\right)\simeq0.6M_{f}$ and $M_{f}\left(3r_{s}^{*}\right)\simeq0.9M_{f}$,
proving the previous claims about the matter distribution. To conclude
the calculation, we must determine $\alpha$ and $\beta$. From
$M_{f}+M_{BH}=\alpha M_{BH}$ it is immediate that
\begin{equation}
\alpha={1 \over 1-{8\beta \over 15}}.
\label{eq:alphabeta}
\end{equation}
The remaining constant, $\beta$, can be fitted using data from the
exact solution, yielding the result $\beta\simeq1.28$. Fig (2) illustrates
a graphic comparison between the exact solution and the approximate
one. The difference between the two never exceeds 7\% of the total
mass. 
\begin{figure}

\centering

\includegraphics{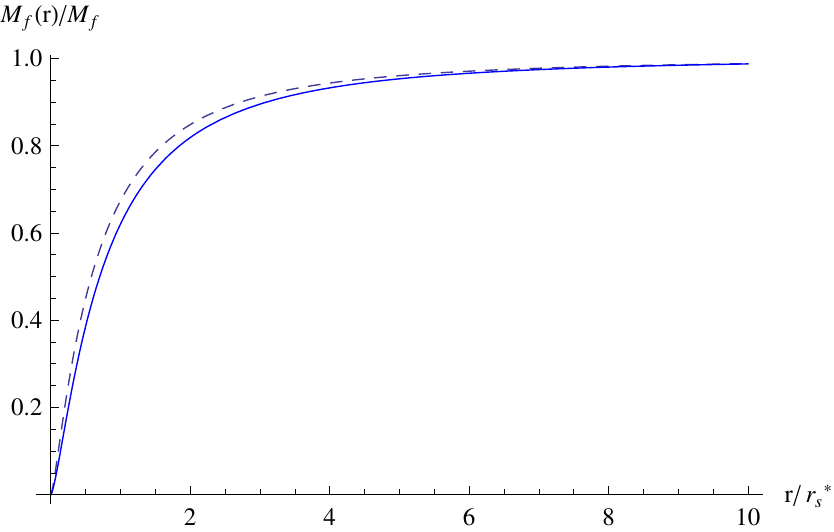}

\caption{Comparison between the exact $M_{f}\left(r\right)$ solution (dashed)
and the approximate one (full) for $\beta=1.28$. The difference between
the two never exceeds 7\% of the total mass.}

\end{figure}
Using this result, we finally obtain
\begin{eqnarray}
\alpha&\simeq&3.15, \nonumber \\
M_{f}&\simeq&2.15M_{BH}. \label{eq:alphaMf}
\end{eqnarray}
Eq. (\ref{eq:alphaMf}) predicts that the total fluid mass in the
vicinity of a spherical, static black hole should be roughly twice
that of the black hole itself. Notice that if one had used the general
relativistic equation $R=\rho/2\kappa$, the condition on $\alpha$
would be (assuming $\beta\simeq1$) $\alpha=1+\alpha$. Although this
equation has no solution, one can think that in the limit $\alpha\gg1$
the equality is approximately verified, leading to $M_{f}^{GR}\gg M_{BH}^{GR}$,
and, consequently, $M_{f}^{GR}\gg M_{f}^{NMC}$, suggesting, as previously
remarked, that the NMC's space-time accommodates much less matter in equilibrium
than its general relativistic counterpart. This strengthens the idea
that the NMC acts as an effective fluid pressure.

\section{Generalization of the Newtonian potential for a perfect fluid}
\label{sec: GNP}

Consider the non-conservation Eq. (\ref{eq:NCT}) and the stress-energy
tensor of a perfect fluid 
\begin{equation}
T^{\mu\nu}=\left(\rho+p\right)u^{\mu}u^{\nu}+pg^{\mu\nu}.
\label{eq:TroP}
\end{equation}
Introducing the projection tensor $h^{\mu\nu}=g^{\mu\nu}+u^{\mu}u^{\nu}$,
Eq. (\ref{eq:NCT}) can be used to show that the equation of motion for
a fluid element is given by~\cite{ExF}
\begin{equation}
{du^{\mu} \over d\tau}+\Gamma_{\alpha\beta}^{\mu}u^{\alpha}u^{\beta}=f^{\mu},
\label{eq:EqMotion}
\end{equation}
where the RHS reads
\begin{equation}
f^{\mu}={1 \over p+\rho}\left[{\lambda F_{2} \over 1+\lambda f_{2}}\left(\mathcal{L}_{m}-p\right)\nabla_{\nu}R-\nabla_{\nu}p\right]h^{\mu\nu}.
\label{eq:Force}
\end{equation}
Regarding Eq. (\ref{eq:Force}), the second term on the right is just
the usual pressure gradient acceleration. However, the first is an
extra force that arises from the NMC. If one inserts $\mathcal{L}_{m}=-\rho$
and uses the previously proven identity~%
\footnote{In fact, the result was proven only for $R\equiv R\left(r,\theta\right)$.
However, its generalization to the the case $R\equiv R\left(x^{0},x^{1},x^{2},x^{3}\right)$
is straightforward. %
} $\lambda F_{2}\nabla_{\nu}R=\partial_{\nu}\left(1+\lambda f_{2}\right)=\partial_{\nu}\left(\lambda f_{2}\right)$,
it can be cast in the form
\begin{equation}
f_{NMC}^{\mu}=-{\partial_{\nu}\left(\lambda f_{2}\right) \over 1+\lambda f_{2}}h^{\mu\nu}.
\label{eq:ExForce}
\end{equation}
Once again, multiplying numerator and denominator by $\rho$ and assuming
a near constant density, one ends up with a NMC force that resembles
the pressure gradient case of GR.

Assuming the usual Newtonian limit approximations $g_{\mu\nu}=\eta_{\mu\nu}+\epsilon_{\mu\nu}$,
where $\eta_{\mu\nu}=diag\left(-1,+1,+1,+1\right)$ and $\left|\epsilon_{\mu\nu}\right|\ll1$,
$\frac{dt}{d\tau}\simeq1$ and $\left|\frac{dx^{i}}{dt}\right|\ll1$,
the LHS of Eq. (\ref{eq:EqMotion}) can be simplified to
\begin{equation}
{du^{i} \over d\tau}+\Gamma_{\alpha\beta}^{i}u^{\alpha}u^{\beta}\simeq{d^{2}x^{i} \over dt^{2}}-{1 \over 2}\partial_{i}\epsilon_{tt},
\label{eq:EqMSimpI}
\end{equation}
the index $i$ running over the space coordinates. Noting additionally
that $h^{i\nu}\simeq\eta^{i\nu}=\delta^{i\nu}$, the full Eq. (\ref{eq:EqMotion})
reads approximately
\begin{equation}
{d^{2}x^{i} \over dt^{2}}=\partial_{i}\left[{g_{tt}+1 \over 2}-\ln\left(\left|1+\lambda f_{2}\right|\right)\right]-{1 \over \rho+p}\partial_{i}p,
\label{eq:EqMSimpII}
\end{equation}
where the identities $\epsilon_{tt}=g_{tt}+1$ and $\partial_{i}\left(\lambda f_{2}\right)/\left(1+\lambda f_{2}\right)=\partial_{i}\ln\left(\left|1+\lambda f_{2}\right|\right)$
were used. It can be seen that the NMC gives rise to an additional force,
and its form suggests the gravitational potential should be generalized
to
\begin{equation}
\Phi=\ln\left(\left|1+\lambda f_{2}\right|\right)-{1 \over 2}\left(g_{tt}+1\right)+\Phi_{0}.
\label{eq:NewPot}
\end{equation}
The inclusion of an extra term in the potential can lead to a vast
number of consequences~\cite{NPotI,NPotII}. As a practical example, it is not difficult
to envisage a coupling, whose effects would only be noticeable at
a galactic level, that could account for the flattening of
the rotation curves of galaxies~\cite{DM}. However, such study is clearly outside
the scope of this work. Instead, one may try to further explore the
physical meaning of a NMC in a rather general context. To achieve
that, consider the usual Newtonian choice $g_{tt}+1\simeq\frac{r_{sc}}{r}$,
where $r_{sc}$ is a constant that sets the length scale (typically
the Schwarzschild radius), and a coupling of the form
\begin{equation}
1+\lambda f_{2}^{\pm}=1\pm\left({r_{sc} \over r+r_{sc}}\right)^{n}.
\label{eq:NMCExample}
\end{equation}
The different signs are used to study both $1+\lambda f_{2}>1$ and
$1+\lambda f_{2}<1$ cases. It is also convenient to take $n>1$ so that
the term $\frac{1}{2}\left(g_{tt}+1\right)$ dominates when $r\rightarrow+\infty$,
although other choices may also be admissible, and indeed, we shall
show that the case presented in section \ref{subsec: A} yields a physically meaningful
coupling that may be cast in the form of Eq. (\ref{eq:NMCExample})
with $n=1$ when $r\gg r_{sc}$. Defining $f_{r}^{GR}=-r_{sc}/2r^{2}$
and neglecting the pressure gradient, the radial force per unit mass
$f_{r}^{\pm}$ reads
\begin{equation}
f_{r}^{\pm}=f_{r}^{GR}+{n \over r+r_{sc}}{\left({r_{sc} \over r+r_{sc}}\right)^{n} \over \left({r_{sc} \over r+r_{sc}}\right)^{n}\pm1}.
\label{eq:RadForce}
\end{equation}
Notice the last term is positive (negative) if one takes the $+$ ($-$) sign. This leads to
\begin{eqnarray}
\left|f_{r}^{+}\right|&<&\left|f_{r}^{GR}\right|, \nonumber \\
\left|f_{r}^{-}\right|&>&\left|f_{r}^{GR}\right|, \label{eq:f+/-}
\end{eqnarray}
which tells us that the corrected gravitational force is weaker (stronger)
than the Newtonian one when $1+\lambda f_{2}>1\,\,\left(1+\lambda f_{2}<1\right)$.
Fig. (3) exhibits this feature graphically for $n=3$. Notice that
Eqs. (\ref{eq:f+/-}) are valid even if $0<n\leq1$. This result reinforces
the arguments already presented in section 2: it seems that a supra-unitary
coupling weakens the link between geometry and matter, while a sub-unitary
one strengthens that connection.

To conclude, let us study the application of the generalized gravitational
potential derived above to the situation described in the first part
of this paper.

Inserting Eq. (\ref{eq:CouplegttRelation}) into Eq. (\ref{eq:NewPot}), one
has
\begin{equation}
\Phi=-{1 \over 2}\left[g_{tt}+1+\ln\left(-g_{tt}\right)\right].
\label{eq:NewPotEx}
\end{equation}
If $g_{tt}$ is written as $g_{tt}=-1+\xi\left(r\right)$, where $\xi\left(r\right)$
is a positive decreasing function of $r$ obeying $\xi\left(r\right)<1$
(a condition imposed by $g_{tt}<0$), then
\begin{equation}
\partial_{r}\Phi={\xi'\left(r\right) \over 2}{\xi\left(r\right) \over 1-\xi\left(r\right)}.
\label{eq:DPot}
\end{equation}

\begin{figure}

\centering

\includegraphics{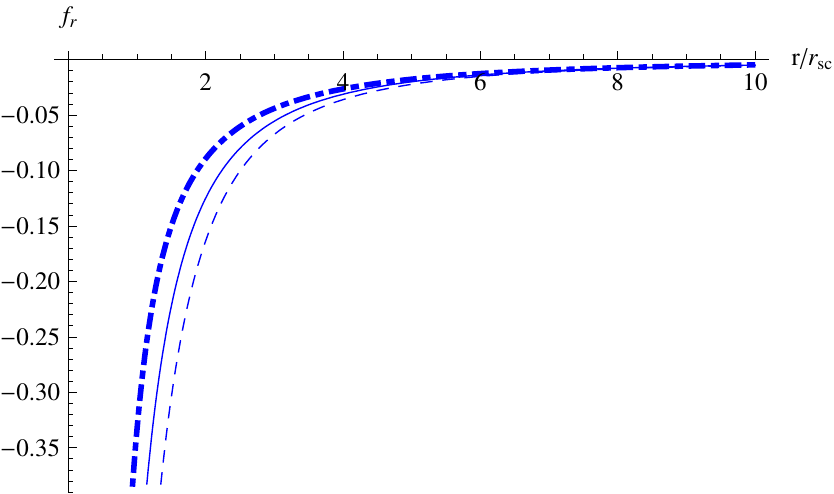}

\caption{Comparison between $f_{r}^{GR}$ (full), $f_{r}^{+}$ (dotdashed)
and $f_{r}^{-}$ (dashed) for $n=3$. It is visible that $\left|f_{r}^{+}\right|<\left|f_{r}^{GR}\right|$,
meaning the corrected gravitational force is slightly weaker than
the Newtonian one when $1+\lambda f_{2}>1$. By the same token, a
coupling with $1+\lambda f_{2}<1$ yields a stronger gravitational
pull. Once again, this suggests that a supra-unitary couple is disfavorable
(in the sense that it weakens the connection between geometry and
matter), while a sub-unitary one is favorable.}

\end{figure}

This gives rise to a positive (repulsive) radial force. To understand
this result, recall, firstly, that we are dealing with a situation
of equilibrium. Indeed, using Eqs. (\ref{eq:gttCpl}) and (\ref{eq:grr})
one can easily verify that Eq. (\ref{eq:EqMotion}) is exactly satisfied
when $u^{\mu}=\left(u^{t},0,0,0\right)$ with $\left(u^{t}\right)^{2}=-\left(g_{tt}\right)^{-1}$.
However, if this equilibrium is disturbed by the addition of a fluid
element, for example, that element is repelled to the outskirts. This
is in agreement with the interpretation of the NMC as an effective
pressure. In a general relativistic non-vacuum static situation, the
gravitational pull is balanced by the pressure gradient to avoid collapse,
and a similar situation occurs: the fluid particle will be attracted
through vacuum, but once it steps inside the matter permeated region
the pressure will exert a repulsive force that pushes the element
outwards. Since one is assuming space is completely filled up with
matter (i.e., $\rho\left(r\right)$ reaches out to infinity), the
force experienced by a test fluid element is always repulsive. Naturally,
in a more refined and realistic model one could assume that matter
is concentrated within a spherical shell, with vacuum on the outside.
Nevertheless, the example considered is sufficient to convincingly suggest
that the effects of a non-minimal coupling mimic those of an effective
pressure.

\section{Conclusion}
\label{sec: Conc}

In this work, we have explored the consequences of a modified theory of gravity with a non-minimal coupling
between geometry and matter as suggested by action Eq. (\ref{eq:S}). In order to do so, a general axially symmetric,
static situation was considered in first place, assuming a matter
distribution modeled by a pressureless perfect fluid. This led to
Eq. (\ref{eq:gRRelation}), and, ultimately, to Eq. (\ref{eq:CouplegttRelation}),
which shows that the coupling, $1+\lambda f_{2}$, and $g_{tt}$ are closely
related through a rather simple functional form.

The study of a spherically symmetric black hole surrounded by pressureless
matter was then examined. Since an exact Schwarzschild solution could
not be considered due to restrictions on $g_{tt}$, one assumed metric
coefficients that resembled the vacuum case when $r\rightarrow+\infty$,
justified by admitting that sufficiently far from the black hole one
should have an {}``almost Schwarzschild'' solution, while being
significantly different in the inner regions, that is, $r\sim r_{s}$.
This permitted an explicit calculation of the matter density $\rho$,
which was shown to approximately fit into the generalized spherical
cusped profile widely present in the literature. The total mass of
fluid was then determined and shown to be about twice the mass of
the black hole. A brief comparison with the general relativistic case,
which predicted a much larger amount of mass, suggested that the non-minimal
coupling could be regarded as an effective pressure, and stipulated
the conditions under which this coupling could be regarded as favorable
or unfavorable to the connection between geometry and matter: $1+\lambda f_{2}<1$
and $1+\lambda f_{2}>1$, respectively.

Finally, in section \ref{sec: GNP} it was demonstrated that when one considers the Newtonian
limit in a non-minimal coupling theory, an additional logarithmic
term appears in the classical gravitational potential. It seems rather
clear that such modification may have a wide range of implications.
At the very least, it can be used to place constraints on the non-minimal
coupling through comparison with Solar System tests~\cite{f2rI}. However, such
procedure would be well outside the scope of this paper, and, as such, will be pursued elsewhere. Instead, 
an argument was presented which solidifies
the idea that a supra-unitary coupling weakens the association between
gravity and matter, whereas a sub-unitary one strengths that relation.
To conclude, the case presented in section \ref{sec: SAS} was inspected
under the modified gravitational potential, leading, once again, to
the idea that the non-minimal coupling plays the role of an effective
pressure.

Although not particularly focused on any fracturing issue of contemporary
cosmology (the accelerated expansion of the Universe~\cite{AE}, the
flattening of the rotation curves of galaxies~\cite{DM}, or the mimicking of the cosmological constant~\cite{CC}, which were previously 
considered in the context of NMC theories), the present work shows that a theory with a non-minimal coupling between curvature
and matter can yield some new and physically
relevant results, especially when applied to astronomical objects
whose nature is still poorly known, such as black holes.


\begin{thebibliography}{99}

\bibitem{fr} T. P. Sotiriou and V. Faraoni, {\it Rev. Mod. Phys.} \textbf{82}, 451 (2010).

\bibitem{ExF} O. Bertolami, C. G. Boehmer, T. Harko and F. S. N. Lobo, {\it Phys. Rev.} \textbf{D 75}, 104016 (2007).

\bibitem{f2rII} O. Bertolami, T. Harko, F. S. N. Lobo and J. P\'aramos, arXiv:0811.2876 [gr-qc].

\bibitem{f2rI} O. Bertolami and J. P\'aramos, {\it Class. Quant. Grav.} \textbf{25}, 245017 (2008).

\bibitem{Lm} O. Bertolami, F. S. N. Lobo and J. P\'aramos, {\it Phys. Rev.} \textbf{D 78}, 064036 (2008).

\bibitem{SS} S. Weinberg, {\it Gravitation and Cosmology: Principles and Applications of the General Theory of Relativity}, John Wiley \& Sons, Inc. (1972).

\bibitem{GCPI} P. Salucci and M. Persic, {\it ASP Conference Series} \textbf{117}, 1 (1997).

\bibitem{GCPII} D. Merritt, J. F. Navarro, A. Ludlow and A. Jenkins, {\it The Astrophysical Journal} \textbf{624}, L85 (2005).

\bibitem{GCPIII} D. Merritt, A. Graham, B. Moore, J. Diemand and B. Terzic, {\it The Astronomical Journal} \textbf{132}, 62685 (2006).

\bibitem{GCPIV} G. van de Ven, R. Mandelbaum and C. R. Keeton, {\it Mon. Not. Roy. Astro. Soc.} \textbf{398}, 607 (2009).

\bibitem{NPotI} S. Capozziello, V. F. Cardone and A. Troisi, {\it Mon. Not. Roy. Astro. Soc.} \textbf{375}, 1423 (2007).

\bibitem{NPotII} C. G. Boehmer, T. Harko and F. S. N. Lobo, {\it Astroparticle Physics} \textbf{29}, 386 (2008).

\bibitem{DM} O. Bertolami and J. P\'aramos, {\it JCAP} \textbf{1003}, 009 (2010).

\bibitem{AE} O. Bertolami, P. Fraz\~ao and J. P\'aramos, {\it Phys. Rev.} \textbf{D 81}, 104046 (2010).

\bibitem{CC} O. Bertolami and J. P\'aramos, arXiv:1107.0225 [gr-qc], to appear in {\it Phys. Rev.} \textbf{D}.






\end{thebibliography}
\end{document}